


\input jytex
\typesize=12pt       \baselinestretch=1200

\def\symbols#1{\ifcase#1 $\bullet$\err@badcountervalue
	\or*\or\markup\dag\or\markup\ddag\or\markup\S%
	\or**\or\markup{\dag\dag}\or\markup{\ddag\ddag}\or\markup{\S\S}%
	\else	$\bullet$\err@badcountervalue
	\fi}
\footnotenumstyle{symbols}
\newcount\tabnum  \tabnum=0
\def\tabnumstyle#1{\global\expandafter\let\expandafter
	\tabnumtype\csname#1\endcsname}
\def\tablabel#1{\global\advance\tabnum by1
  \label{#1}{\hbox{\tabnumtype\tabnum}}\putlab{#1}}
\def\Table#1{\hbox{Table \tablabel{#1}}}
\def\puttab#1#2#3{\goodbreak \midinsert #2 \bigskip
  \centertext{\bf Table \putlab{#1}: #3} \endinsert}
\tabnumstyle{arabic}

\sectionstyle{center}  \sectionnumstyle{arabic}
\subsectionstyle{left}  \subsectionnumstyle{blank}
\def\sec#1{{\bigfonts\bf\section{#1}}}  \def\subsec#1{{\bf\subsection{#1}}}
\def\eq#1{\eqno\eqnlabel{#1}$$}  \def\ref#1{\markup{[\putref{#1}]}}
\def\puteq#1{Eq.~(\puteqn{#1})}
\def\to{\!\rightarrow\!}  \def\ord#1{{\cal O}(#1)}
\def\Im{{\rm Im}}	  	  
\def\etal{{\it et.al.}}	  \def\ie{{\it i.e.}}	  \def\eg{{\it e.g.}}
\def\kr{k_r}		  \def\kt{k_t}
\def\d{{\rm d}}		  \def\k{{\bf k}}	  \def\x{{\bf x}}
\def\a{\alpha}		  \def\b{\kappa}	  \def\o{\omega}
\def\p{\phi}		  \def\bp{\bar\p}	  \def\del{\partial}
\def\det{{\rm det}}	  	  \def\fct{F^{\rm ct}}
\def\dfo{\Delta F_1}	  \def\dft{\Delta F_T}	  \def\dff{\Delta F_{1+T}}
\def\sech{{\rm sech}}     \def\tanh{{\rm tanh}}	  \def\Ineg{I^{(neg)}}
\def\-{\!-\!}             \def\0{\hfil 0}
\def\pn{\phantom{-}}	  \def\pdigit{\phantom{0}}
\def\frac#1#2{{\textstyle {#1 \over #2}}}
\def\mfrac#1#2{{\textstyle {- \atop }\!{#1 \over #2}}}
\def\pmb#1{\setbox0=\hbox{#1}\kern-.01em\copy0\kern-\wd0
  \kern.02em\copy0\kern-\wd0\kern-.01em\raise.017em\box0}


\pagenumstyle{blank}
\hbox to\hsize{\footnotesize\baselineskip=12pt
  \hfil\vtop{\hbox{\strut CALT-68-1881}
  \hbox{\strut hep-ph/9311353} \hbox{\strut}
  \hbox{\strut DOE RESEARCH AND} \hbox{\strut DEVELOPMENT REPORT}}}
\vskip1.in
\centertext{\baselinestretch=1000\bigsize\bf
The Exact Critical Bubble Free Energy and the\\Effectiveness of Effective
Potential Approximations%
\footnote{Work supported in part by the U.S. Dept. of Energy under Contract
  No. DE-FG03-92-ER40701.}}
\vskip .25in \footnotenumstyle{arabic}
\centertext{David E. Brahm%
\footnote[1]{Email: \tt brahm@fermi.phys.cmu.edu}
\\{\it Carnegie Mellon, Pittsburgh PA 15213}
\\and
\\Clarence L. Y. Lee%
\footnote[2]{Email: \tt lee@theory3.caltech.edu}
\\{\it California Institute of Technology, Pasadena, CA 91125}}
\vskip .2in \footnotenumstyle{symbols}

\vskip .7in
\centerline{\bf Abstract}
\medskip {\narrower \baselinestretch=1100
To calculate the temperature at which a first-order cosmological phase
transition occurs, one must calculate $F_c(T)$, the free energy of a critical
bubble configuration.  $F_c(T)$ is often approximated by the classical energy
plus an integral over the bubble of the effective potential; one must choose a
method for calculating the effective potential when $V''<0$.  We test
different effective potential approximations at one loop.  The agreement
is best if one pulls a factor of $\mu^4/T^4$ into the decay rate prefactor
[where $\mu^2 = V''(\p_f)$], and takes the real part of the effective
potential in the region $V''<0$.  We perform a similar analysis on the
1-dimensional kink. \par}

\vfill\lefttext{November, 1993}\medskip
\newpage \pagenum=0 \pagenumstyle{arabic}

\sec{Introduction}
\subsec{Thermal Tunneling and the Critical Bubble Free Energy}

A scalar field theory whose potential $V$ has two local minima may tunnel out
of the false vacuum ($\p_f$) by the nucleation and subsequent growth of bubbles
of true vacuum ($\p_t$).  While we will refer to $V$ as the ``classical''
potential, it may arise in part from integrating out other particles in the
theory, \eg\ gauge bosons\ref{ejw}, so $V$ may have implicit temperature ($T$)
dependence.  The nucleation rate per unit volume in the static limit ($RT \gg
1$) is calculated in the Gaussian approximation (\ie\ to 1-loop order) to
be\ref{cc,affleck,fTref}
$${\Gamma\over{\cal V}} = {1\over{\cal V}} \, {|\o_-|\over\pi T} \, {1\over2}
  \, T \left| \det[\del^2 + V''(\bp)] \over \det[\del^2 + \mu^2]
  \right|^{-1/2} e^{-E_c/T} \eq{g1}
where $\mu^2 = V''(\p_f)$.  $E_c$ is the classical energy of the critical
bubble, a static and spherically symmetric field configuration $\bp(r)$, of
radius $R$, which extremizes the classical action\ref{gleiser} subject to
periodic boundary conditions in Euclidean time.  The determinants range over a
complete basis of fluctuations about the classical solution ($\bp(r)$ or
$\p_f$), subject to the same boundary conditions.  $\o_-^2 < 0$ is the
eigenvalue of the ``breathing'' mode about $\bp(r)$.  The second term on the
RHS of \puteq{g1} is from Affleck\ref{affleck}, and the $\frac12$ is from
analytically continuing the breathing mode integration\ref{cc}.

With the periodic boundary conditions,
$$\det[\del^2 + V''(\bp)] = \exp\left\{ \sum_{n=-\infty}^\infty \sum_j \ln
  \left[ (2\pi n T)^2 + \o_j^2 \right] \right\} \;,\eq{trln}
where the $\o_j^2$ are eigenvalues of $[-\nabla^2 + V''(\bp)]$, and the
$(\o_j^0)^2$ are eigenvalues of $[-\nabla^2 + \mu^2]$.  We use the
identity\ref{dj}
$$\frac{T}2 \sum_n \ln\left[ (2\pi n T)^2 + \o^2 \right] =
  \frac\o2 + T \ln(1-e^{-\o/T}) + {\cal C} =
  T \ln\left[ 2\,\sinh\left(\frac\o{2T}\right) \right] + {\cal C} \;.\eq{ident}
The constants $\cal C$ cancel out in \puteq{g1}.  The $\o_-$ contribution is
then traditionally pulled back into the prefactor.  The 3 ``translation''
modes ($n\!=\!0$ and $\o_0\!=\!0$) are not treated correctly above; they
actually give ${\cal V} (E_c/2\pi T)^{3/2}$ in the prefactor\ref{cc}, and the
remaining $\o_0$ contribution (from $n\!\ne\!0$ modes) gives $T^3$ in the
prefactor.  This gives
$${\Gamma\over{\cal V}} = {T^4\over2\pi} \left(E_c\over 2\pi T \right)^{3
  /2} {|\o_-|/2T \over \sin(|\o_-|/2T)} \; e^{-F_c^{trad}/T} \,\eq{g2}
where the ``traditional'' bubble free energy
$$F_c^{trad} \;\equiv\; E_c + \dff^{trad} \;\equiv\; E_c + \dfo^{trad} +
  \dft^{trad} \;,\eq{deco}
$$\dfo^{trad} = {\sum_j}' {\o_j-\o_j^0 \over 2} + \fct, \qquad \dft^{trad} =
  {\sum_j}' T \ln\left( 1 - e^{-\o_j/T} \over 1 - e^{-\o_j^0/T} \right)
  \;.\eq{dfdef}
Primes on the sums in \puteq{dfdef} indicate omission of the translation and
breathing modes ($\o_j,\; j=1\-4$).  Counterterms $\fct$ are discussed below.

We now define\footnote{This is somewhat like removing the lowest 4 $\o_j^0$'s
from the sums in \puteq{dfdef}, in addition to the lowest 4 $\o_j$'s, since
their contribution to $F_c^{trad}$ is $-4 [\frac\mu2 + T \ln(1-e^{-\mu/T})]
\approx 4 T \ln(T/\mu)$.}
$$F_c^{sub} \;\equiv\; E_c + \dff^{sub} \;\equiv\; E_c + \dfo^{sub} +
  \dft^{sub} \;,\eq{Fcdef}
$$\dfo^{sub} \equiv \dfo^{trad}, \qquad \dft^{sub} \equiv \dft^{trad} -
  4T\ln(T/\mu) \;.\eq{subdef}
Now \puteq{g2} becomes
$${\Gamma\over{\cal V}} = {\mu^4\over2\pi} \left(E_c\over 2\pi T \right)^{3
  /2} {|\o_-|/2T \over \sin(|\o_-|/2T)} \; e^{-F_c^{sub}/T} \;.\eq{g3}
We will find that the effective potential approximation most closely
approximates $F_c^{sub}$.

\subsec{The Effective Potential}

The sums in \puteq{dfdef} are often approximated by treating the fluctuations
locally as plane waves to get an effective potential $V_{1+T} = V_1 + V_T$,
then integrating $[V_{1+T}(\bp) - V_{1+T}(\p_f)]$ over all space.  No attempt
is made to remove the 4 translation and breathing modes.  In \puteq{dfdef}
one substitutes
$$\sum_j \to \int\d^3 \x \int_0^\Lambda {\d^3 \k\over (2\pi)^3},
  \qquad \omega_j \to \sqrt{\k^2+V''(\bp)}, \quad
  \omega_j^0 \to \sqrt{\k^2+\mu^2} \;,\eq{trk}
and one finds, with $m^2 \equiv V''(\p)$,
$$V_1(\p) = {1\over64\pi^2} \left\{ m^4 \ln\left( \frac{m^2}{\mu^2} \right) -
  \frac32 m^4 + 2 m^2 \mu^2 - \frac12 \mu^4 \right\} \;,\eq{V1}
$$V_T(\p) = {T^4 \over 2\pi^2}\; I(m/T),
  \qquad I(y) \equiv \int_0^\infty \d x \> x^2
  \ln\left( 1 - e^{-\sqrt{x^2+y^2}} \right) \;.\eq{VT}
The expansion of $I(y)$ for real $y<2\pi$ is\ref{dj,haber}
$$I(y) = {-\pi^4\over45} + {\pi^2\over12}y^2 - {\pi\over6}y^3
  - {y^4\over32} \left[ \ln y^2 - c_3 + \sum_{k=1}^\infty {4 (2k)!\,
  \zeta(2k\!+\!1) \over k! (k\!+\!2)!} \left(-y^2\over16\pi^2\right)^k
  \right]  \eq{Iexp}
where $c_3 = \frac32 + 2\ln(4\pi)-2\gamma \approx 5.4076$.  We choose a
renormalization scheme in which all divergent graphs are precisely
cancelled by counterterms so that at zero external momenta,
$V_1(\p_f) = V'_1(\p_f) = V''_1(\p_f) =0$ (and there is no
wavefunction renormalization),\ref{clee} specifically:
$$\fct = {-1\over64\pi^2} \int\d^3 \x \left. \left\{ \left[4\Lambda^4 + \frac12
  \mu^4 \right] + m^2 [4\Lambda^2-2\mu^2] + m^4 \left[ 2 - \ln\left(
  \frac{4\Lambda^2}{\mu^2} \right) \right] \right\}
  \right|_{m^2=\mu^2}^{m^2=V''} \;.\eq{fct}

In the region $m^2<0$, we must modify these results to give a real answer.
For $V_1$ we will always take the real part of \puteq{V1}.  For $V_T$ let us
keep the first equation of \puteq{VT}, but replace $I(m/T)$ by
$\Ineg(|m|/T)$ where
$$\Ineg(Y) \equiv {-\pi^4\over45} - {\pi^2\over12}Y^2 + Y^3 \left[a +
  b\ln(Y^2)\right] - {Y^4\over32} \left[\ln(Y^2) - c_3 + c\right] + \cdots
  \;.\eq{Ineg}
Methods we consider are then parametrized by $\{a,b,c\}$.  The most common and
obvious method (A) is to take the real part of \puteq{Iexp}, corresponding to
$\{a=b=c=0\}$.  Another method (B), proposed in ref.~[\putref{us}], replaces
the lower limit of integration in \puteq{trk} by $k=\Im\{m\}$ (eliminating
fluctuations with wavelengths longer than the bubble thickness), and
corresponds to $\{a=\frac49-\frac13 \ln(2),\, b=\frac16,\, c=0\}$.

\subsec{The Derivative Expansion}

For configurations $\bp(\x)$ which vary slowly, the effective potential
approximation is the leading term in a derivative expansion of the free
energy. The next term (at high $T$) is\ref{chan,moss}
$$\dft^{der} - \dft^{pot} = {T\over 192\pi} \int\d^3 \x \, m^{-1} \,
  \nabla^2(m^2) \;,\eq{fd2}
and again we take the real part (Method A) when necessary.  More terms are
given explicitly in ref.~[\putref{moss}]; they become increasingly divergent at
$m^2=0$, where the derivation breaks down (because an integration by parts
becomes invalid).  Also, no attempt is made to omit modes.  The usefulness of
\puteq{fd2} is thus highly suspect, but we note that derivative corrections are
predicted to be $\ord{T^1}$.

\subsec{Scales, Approximations, and Goals}

Our generic tree-level potential will be quartic in $\p$ with $\p_f=0$,
$V''(0)=\mu^2$, and $\p_t=\sigma$.  By rescaling\ref{thick} $\p=\sigma
\tilde\p$, $x=\tilde x/\mu$, and $T=\mu\tilde T$, we can rewrite the
4-action $S_0$ as
$$S_0 = {E_c\over T} = \left( \sigma\over\mu \right)^2 {1\over\tilde T}
  \int\d^3\tilde\x\, \left\{ {1\over2} \left(\d\tilde\p\over\d\tilde r
  \right)^2 - \left[{1\over2}\tilde\p^2 - {2\b+1\over3}\tilde\p^3 +
  {\b\over2}\tilde\p^4 \right] \right\} \;.\eq{scale}
$\b\ge1$ is a dimensionless parameter; $\b\to 1$ (degenerate minima) is the
thin-wall limit, while larger $\b$ gives thicker bubbles.  With tildes
indicating dimensionless results,
$$E_c = (\sigma^2/\mu)\,\tilde E_c, \qquad \Delta F_{1,T} =
  \mu\,\Delta\tilde F_{1,T} \;.\eq{tdefs}
The loop expansion\ref{cnl} is an expansion in $(\mu/\sigma)^2$ and $\tilde T$.
It is sometimes claimed that higher loops should eliminate the complex terms in
$F_c^{pot}$, but this cannot be generally true since the higher-loop
contributions are suppressed by these arbitrary parameters.  Henceforth we will
drop the tildes and work in the rescaled theory (\ie\ set $\mu=\sigma=1$).

We always use the static approximation\ref{alpriv} ($RT \gg 1$) and the 1-loop
approximation.  In Section 3 we will use the thin-wall approximation, $R \gg
1$.  At times we will make high-temperature expansions, requiring $T \ge 1$
(note the thin-wall and high-temperature limits together imply the static
limit).  We are examining the validity of the effective potential
approximation.

In this paper we will study several systems: the 1-dimensional kink, the
thin-wall bubble, and two thick-wall bubbles.  We will calculate $\dfo$ and
$\dft$ for each system exactly [$F_c^{sub}$ in \puteq{subdef}], in the
effective potential approximation [$F_c^{pot}$ from
Eqs.~(\puteqn{V1}-\puteqn{VT}), using different methods to calculate
$\Ineg$ in \puteq{Ineg}], and using the next term of the derivative
expansion [$F_c^{der}$ from \puteq{fd2}].

\sec{The 1-Dimensional Kink}
\subsec{Classical Results}

We warm up by calculating the free energy of a kink in 1 spatial
dimension:\ref{dod}
$${\d^2 \bp \over \d x^2} = V'(\bp), \qquad
  {\d \bp \over \d x} = -\sqrt{2V(\bp)}, \qquad
  V(\p) = \frac12 \p^2 (1-\p)^2 \;.\eq{cb2}
The potential is that of \puteq{scale} with $\b=1$.
The kink solution is (up to an arbitrary shift in coordinate)
$$\bp(x) = \frac12 [1 - \tanh(\frac12 x)], \qquad
  V''(\bp(x)) = 1 - \frac32 \sech^2(\frac12 x) \;.\eq{kink}
\puteq{cb2} allows us to convert integrals over $x$ into integrals
over $\p$:
$$\int_{-\infty}^\infty \d x \;\to\; \int_0^1 {\d\p \over \p(1\-\p)}
  \;.\eq{xtoq}
For example, the classical energy is
$$E_c^{1D} = \int_0^1 {\d\p \over \p(1\-\p)} \p^2(1\-\p)^2 = {1\over6}
  \;.\eq{s1}
Note that in 1D [compare to \puteq{tdefs}] $E_c = \mu\sigma^2 \,\tilde E_c$
and $\Delta F_{1,T} = \mu\,\Delta\tilde F_{1,T}$, so with scales restored
$E_c^{1D} = \mu\sigma^2/6$.

\subsec{Exact Results from the Eigenvalue Sum}

The solutions to the eigenvalue equations (setting $\mu=1$) are
known:\ref{dod,DHN}
$$\o_s^0 = \sqrt{(k_s^0)^2 + 1}, \qquad \o_1=0, \quad
  \o_2 = \sqrt3/2, \quad \o_{s>2} = \sqrt{(k_s)^2 + 1},$$
$$k_s^0 = {\pi s\over L}, \qquad k_s = {\pi s - \delta(k_s) \over L},
  \quad \delta(k) = 2\pi - 2\tan^{-1}(k) - 2\tan^{-1}(2k) \;,\eq{ksol}
where we have imposed vanishing boundary conditions on a box of length $L$,
so $s$ is a positive integer.  We drop the translation mode eigenvalue
$\o_1$; there is no negative eigenvalue in 1D.  In the continuum limit,
$$\dfo^{trad} = {\sqrt3\over 4} + \int_0^\Lambda {\d k\over\pi}
  {\d\delta \over \d k} {\sqrt{k^2+1}\over2} - {3\over2\pi} + \fct,$$
$$\dft^{trad} = T \ln\left(1 - e^{-\sqrt3/2T}\right) + \int_0^\infty
  {\d k\over\pi} {\d\delta \over \d k} \,T \ln\left(1 - e^{-\sqrt{k^2+1}
  /T} \right) \;.\eq{kdf}
In our renormalization scheme the 1D counterterms analogous to \puteq{fct} are
$$\fct = {-1\over16\pi} \int\d x \left\{ \left[4\Lambda^2 +1\right] + m^2
    \left[2 + 2\ln(4\Lambda^2)\right] - m^4 \right\} \bigr|_{m^2=1}^{m^2=V''}
  = {1\over 8\pi} [3 + 6 \ln(4\Lambda^2)] \;.\eq{fct1d}
(This differs from ref.~[\putref{dod}] by $3/8\pi$ due
to different renormalization schemes; also note their $m^2\equiv \mu^2/2$.)
We define $\dfo^{sub} \equiv \dfo^{trad}$ and $\dft^{sub} \equiv \dft^{trad} -
T\ln(T/\mu)$, and find
$$\dfo^{sub} = {1\over4\sqrt3} - {9\over8\pi} = -.2138 \;.\eq{dodres}
$$\dff^{sub} = - (\ln\sqrt{12})\, T + {3\over2\pi}\ln(T) + {6c_1-3\over8\pi}
  + {3\zeta(3)\over 32\pi^3}\, T^{-2} + \cdots \eq{kftres}
where $c_1 = 1 + 2\ln(4\pi)-2\gamma \approx 4.9076$, and $\zeta(3) \approx
1.2021$.  These results are in the row marked ``sub'' of \Table{tone}.

\puttab{tone}{
 \centerline{ \vbox{\offinterlineskip
 \halign to\hsize{\vrule#\tabskip=.5em plus1em minus.5em&
   \hfil# & \vrule# & $#$\hfil & \vrule# &
   $#$\hfil & $#$\hfil & $#$\hfil & $#$\hfil & $#$\hfil & $#$\hfil &
   \tabskip=0pt\vrule#\cr
 \noalign{\hrule}
 height14pt depth9.5pt& && \hfil\dfo &&\multispan6 \hfil$\dff$\hfil &\cr
 height16pt depth7.5pt& Method && && \hfil T \ln(T) & \hfil
   T & \hfil \ln(T) & \hfil 1 & \hfil T^{-1} & \hfil T^{-2} &\cr
 \noalign{\hrule}
 height14pt depth5.5pt&   sub && -.2138 &&
      \0   & -1.2425 & .4775 & 1.0522 &   \0   & .0036 &\cr
 height12pt depth5.5pt& pot(A) && -.0916 &&
      \0   & -2.1145 & .4775 & 1.0522 &   \0   & .0036 &\cr
 height12pt depth5.5pt& der(A) && -.0916 &&
      \0   & -2.1730 & .4775 & 1.0522 &   \0   & .0036 &\cr
 height12pt depth5.5pt& pot(B) && -.0916 &&
    0.4495 & -1.7222 & .4775 & 1.0522 & .0045  & .0036 &\cr
 \noalign{\hrule}
 } } }
}{Kink free energy in low- and high-T regimes.}

\subsec{1D Effective Potential and Derivative Expansion Results}

The 1D effective potential for real $m$ is\ref{haber}
$$V_1 = {-m^2\over8\pi} \ln(m^2) + {m^4-1 \over 16\pi}, \qquad
  V_T = {T^2\over\pi} \hat I(m/T) \eq{kv}
$$\hat I(y) = {-\pi^2\over 6} + {\pi y\over2} + {y^2\over8} \left[\ln
  (y^2) - c_1 \right] - {\zeta(3) y^4\over 64\pi^2} + \cdots \;.\eq{Ihat}
For $m^2<0$ we replace $\hat I(m/T)$ by $\hat \Ineg(|m|/T)$ where
$$\hat \Ineg(Y) = {-\pi^2\over 6} + Y \left[\hat a + \hat b \ln(Y^2)\right]
  -{Y^2\over8} \left[\ln(Y^2) - c_1 + \hat c\right] + \cdots \eq{Ihneg}
Method A gives $\{\hat a=\hat b=\hat c=0\}$, and Method B gives
$\{\hat a=1-\ln(2),\, \hat b=\mfrac12,\, \hat c=0\}$.

We integrate (the real part of) $V_1$ from \puteq{kv} over all space, using
\puteq{xtoq}, to get $\dfo^{pot(A)} = -.0916$, which differs significantly from
$\dfo^{sub} = -.2138$ (note each result is renormalization-dependent, but the
difference is not).  This difference, which was calculated in
ref.~[\putref{dod}], dominates the low-T regime.

A similar integral for the high-T expansion gives
$$\dff^{pot(A)} = \ln[2(\sqrt3\-\sqrt2)^{\sqrt6}] \,T
  + {3\over2\pi} \ln(T) + {6c_1-3\over8\pi}
  + {3\zeta(3)\over32\pi^3}\, T^{-2} + \cdots \;,\eq{kftd1}
as shown in the line marked ``pot(A)'' of
Table~\putlab{tone}.  Note that the difference between the true result and
the potential approximation no longer lies in the constant term, but only
(as far as we have taken the expansion) in the $T$ term!  It is
$$\dff^{sub}-\dff^{pot(A)} = -\ln\left[ 4\sqrt3 (\sqrt3\-\sqrt2)^{\sqrt6}
  \right]\, T = .8720\, T \;.\eq{kdisc}

The next term of the derivative expansion [analogous to \puteq{fd2}] is
$$\dft^{der} - \dft^{pot} = {T\over 96} \int\d x \, m^{-3} \, \nabla^2(m^2)
  = {\sqrt6\over48} \ln(\sqrt3-\sqrt2) T = -.0585\, T \eq{kdnum}
as incorporated in the third line of Table~\putlab{tone}.  It is a very poor
approximation to \puteq{kdisc}!

Results from Method B are given in the fourth line of Table~\putlab{tone};
these are also unsatisfactory.  In fact, the choice $\{\hat a=1.940,\, \hat
b=\hat c=0\}$ in \puteq{Ihneg} would give the correct (``sub'') results,
but it is not clear if there is any physics in this choice.

\sec{The Thin-Wall Critical Bubble}
\subsec{Classical Results}

For $\b$ close to (but larger than) unity in \puteq{scale}, the solution to
  $$\nabla^2 \bp = V'(\bp) \eq{cb1}
is a thin-wall bubble, given approximately by the kink solution in the
radial coordinate, \puteq{kink} with $x=r-R$ and $R\gg1$.\ref{cc}
The tree-level critical bubble energy has volume and surface terms:
  $$E_c = 4\pi\int r^2 \d r \left[\frac12 \left( \frac{\d\bp}{\d r}
    \right)^2 + V(\bp(r)) \right] \approx -\frac43 \pi R^3 |V(1)| + 4\pi
    R^2 E_c^{1D} \;,\eq{ecnum}
where $E_c^{1D}=\frac16$ was given in \puteq{s1}, and $|V(1)|=(\b\-1)/6$.
We extremize to find the bubble radius $R$ and energy $E_c$,
  $$R = {2\over \b-1}, \qquad E_c = {8\pi\over 9(\b\-1)^2} = {2\pi R^2\over
    9} \;.\eq{RE}
The wall thickness is $\ord{1}$ (\ie\ $\mu^{-1}$).  It can also be
shown\ref{cc} that $\o_-^2 \approx -2/R^2$, so the static and thin-wall
limits imply that the third factor of \puteq{g2} is near unity.

\subsec{Exact Results for a Domain Wall}

In the thin-wall limit, the surface free-energy density $f_{1,T} = \Delta
F_{1,T}/(4\pi R^2)$ of the bubble wall equals that of a planar domain
wall\ref{walls}.  We can thus solve the eigenvalue equation in Cartesian
coordinates, using \puteq{ksol} for the radial wavenumber $\kr$, and plane
waves
for the tangential $\kt$, to get
$$\eqalign{f_1^{sub} =&\; \int_0^\Lambda {\kt\,\d \kt \over 2\pi} \Biggl\{
   {\kt\over2} + {\sqrt{\kt^2+3/4}\over2} - {\sqrt{\Lambda^2+1}\over2\pi}
   \,\delta(\sqrt{\Lambda^2-\kt^2}) \cr
  &\quad + \int_0^{\sqrt{\Lambda^2-\kt^2}} {\d\kr\over\pi} \left( {-2\over
   \kr^2 + 1} + {-4\over 4\kr^2 + 1} \right) {\sqrt{\kt^2+\kr^2+1} \over 2}
   \Biggr\} + {3 \Lambda^2\over8\pi^2} - {3\over32\pi^2} \ln(4\Lambda^2) \cr
  =&\; {-1\over32\pi^2} \left( {\pi\over\sqrt3} + 6 \right) = -.02474 \;, \cr
  f_T^{sub} =&\; T \int_0^\infty {\kt\,\d \kt \over 2\pi} \Biggl\{ \ln
   \left[1 - e^{-\kt/T}\right] + \ln\left[1 - e^{-\sqrt{\kt^2 +
   \frac34}/T}\right] + \int_0^\infty {\d \kr \over \pi} \cr
 &\times \left( {-2\over \kr^2 + 1} + {-4\over 4\kr^2 + 1} \right)
   \ln\left[1 - e^{-\sqrt{\kt^2+\kr^2+1}/T} \right] \Biggr\} \;.\cr} \eq{baz}
We have performed the $f_T$ integral numerically, and fit to an expansion
in $T^{-1}$; the results are shown in \Table{mu/T} in the row marked
``sub''.\footnote{These results are also useful for the study of second-order
phase transitions, in which the domain wall free energy density is set to
zero.\ref{walls}  Restoring units,
$$f[\bp^{wall}] = \mu \left[ {\sigma^2\over6} - {T_c^2\over4} + .15215 \,\mu
  T_c -.01900 \,\mu^2 \ln(T_c/\mu) - \cdots \right] = 0 $$
giving, for $\mu\ll\sigma$, $T_c = \sqrt{2/3} \,\sigma + 0.3 \mu + \cdots$.
That is, the critical temperature is a bit higher than the leading result which
is in the literature.}

\puttab{mu/T}{
 \centerline{ \vbox{\offinterlineskip
 \halign to\hsize{\vrule#\tabskip=.5em plus1em minus.5em&
   \hfil# & \vrule# & $#$\hfil & \vrule# &
   $#$\hfil & $#$\hfil & $#$\hfil & $#$\hfil & $#$\hfil & $#$\hfil & $#$\hfil
   & \tabskip=0pt\vrule#\cr
 \noalign{\hrule}
 height14pt depth7.5pt& && \hfil f_1 &&\multispan7 \hfil$f_{1+T}$\hfil &\cr
 height17pt depth7.5pt& Method && && \hfil T^2 & \hfil T \ln(T) & \hfil T &
   \hfil \ln(T) & \hfil 1 & \hfil T^{-1} & \hfil T^{-2} &\cr
 \noalign{\hrule}
 height14pt depth5.5pt&   sub  && -.02474 && -1/4 &    \0  & .15215 &
   -.01900 & -.03712 &    \0   & -.00012 &\cr
 height12pt depth5.5pt& pot(A) && -.00661 && -1/4 &    \0  & .15452 &
   -.01900 & -.05612 &    \0   & -.00012 &\cr
 height12pt depth5.5pt& der(A) && -.00661 && -1/4 &    \0  & .15187 &
   -.01900 & -.05612 &    \0   & -.00012 &\cr
 height12pt depth5.5pt& pot(B) && -.00661 && -1/4 & .00864 & .16409 &
   -.01900 & -.05612 & .00006  & -.00012 &\cr
 \noalign{\hrule}
 } } }
}{Thin-wall bubble free energy density for low- and high-T.}

\subsec{Effective Potential and Derivative Expansion Results}

Results from integrating the effective potential, and the next term of the
derivative expansion, over the bubble [again using \puteq{xtoq}] are shown
in the rest of Table~\putlab{mu/T}.  Using the general $\Ineg$ of
\puteq{Ineg} gives
$$\eqalign{f_{1+T}^{pot} =& \mfrac14 T^2 - (.0518\,b)\, T \ln(T) + (.1545 +
  .0259\,a - .0242\,b)\, T \cr
  &- (.0190)\,\ln(T) + (-.05612 - .000514\,c) \cr} \;.\eq{eft}
Matching this to the true $f_{1+T}^{sub}$ gives the coefficients $\{a,b,c\}$
shown in the first line ($\b=1$) of \Table{abc}.\footnote{First subtracting the
derivative correction of \puteq{fd2} from $\dff^{sub}$ would give $a$ values of
$.0109$, $.3877$, and $.5128$, respectively.  For the kink it gives
$\hat a=2.070$.  These results are no more enlightening.}

We see ``derivative corrections'' are $\ord{T}$.  The derivative expansion
prediction, $f_{1+T}^{der}$ from \puteq{fd2}, is a reasonable approximation to
them in this case.

\puttab{abc}{
 \centerline{ \vbox{\offinterlineskip
 \halign to3.5in{\vrule#\tabskip=.5em plus1em minus.5em&
   \hfil$#$ & \vrule# & $#$\hfil & $#$\hfil & $#$\hfil &
   \tabskip=0pt\vrule#\cr
 \noalign{\hrule}
 height14pt depth7.5pt& \b\hfil && \hfil a & \hfil b & \hfil c &\cr
 \noalign{\hrule}
 height13pt depth5.5pt&   1 &&    -.0913 & \0 & -36.974 &\cr
 height12pt depth5.5pt& 1.5 && \pn .2834 & \0 & -\pdigit 1.424 &\cr
 height12pt depth5.5pt& 2.5 && \pn .4188 & \0 & -\pdigit 0.180 &\cr
 \noalign{\hrule}
 } } }
}{\pmb{$\Ineg$}\ parameters that make \pmb{$\dff^{pot}=\dff^{sub}$}.}

\sec{Thick-Wall Critical Bubbles}
\subsec{Classical Results}

{}From \puteq{scale}, the (scaled) potential (Fig.~1) is
  $$V = {1\over2}\p^2 - {2\b+1 \over 3} \p^3 + {\b\over2} \p^4 \;.\eq{vthk}
Larger $\b>1$ gives thicker bubbles.  The minima are at $\p=0$ and $\p=1$,
with $V''(0)=1$ and $V''(1)=2\b-1$.  The bubble profile is the
solution to
  $$\bp'' + 2\bp'/r = \bp(1-\bp)(1-2\b \bp) \;.\eq{prok}
Fig.~2 plots $\bp(r)$ and $V''(r)$ for $\b=1.5$ and $\b=2.5$.  From
ref.~[\putref{thick}], the classical energy is approximately
  $$E_c \approx {4.85 \a\over\b} \left[ 1 + {\a\over4} \left(
    1 + {2.4\over 1-\a} + {.26\over(1-\a)^2} \right)\right], \qquad
    \a \equiv {9\b \over (1+2\b)^2} \;.\eq{ecthk}

\subsec{Exact, Effective Potential, and Derivative Expansion Results}

Our method of calculating the exact free energy $F_c^{sub}$, formally given by
\puteq{Fcdef}, is described in ref.~[\putref{clee}].  The results for $\b=1.5$
are in \Table{b1.5}, and for $\b=2.5$ in \Table{b2.5}\footnote{In our fit to
the data, we allowed a $T^{-2}$ term, not shown, and constrained the
$T^2$, $T \ln(T)$, and $\ln(T)$ terms.}, along with effective potential and
derivative expansion approximations.  Thin-wall predictions are also shown for
two values of $R$: one chosen to give the correct $T^2$ coefficient
(``thin-1''), and one given by \puteq{RE} (``thin-2'').  Finally, the
parameters in $\Ineg$ needed to match the effective potential approximation to
the exact result are given in Table~\putlab{abc}.

\puttab{b1.5}{
 \centerline{ \vbox{\offinterlineskip
 \halign to\hsize{\vrule#\tabskip=.5em plus1em minus.5em&
   \hfil# & \vrule# & $#$\hfil & \vrule# &
   $#$\hfil & $#$\hfil & $#$\hfil & $#$\hfil & $#$\hfil
   & \tabskip=0pt\vrule#\cr
 \noalign{\hrule}
 height14pt depth7.5pt& && \hfil\dfo &&\multispan5 \hfil$\dff$\hfil &\cr
 height17pt depth7.5pt& Method && && \hfil T^2 & \hfil T \ln(T) & \hfil T &
   \hfil \ln(T) & \hfil 1 &\cr
 \noalign{\hrule}
 height14pt depth5.5pt&   sub  && -2.13 && -78.61 &        \0  & 49.52 &
  -5.193 & -15.64  &\cr
 height12pt depth5.5pt& pot(A) && -2.65 && -78.61 &        \0  & 45.47 &
  -5.193 & -16.12  &\cr
 height12pt depth5.5pt& der(A) && -2.65 && -78.61 &        \0  & 43.98 &
  -5.193 & -16.12  &\cr
 height12pt depth5.5pt& pot(B) && -2.65 && -78.61 & \hfil 4.76 & 49.73 &
  -5.193 & -16.12  &\cr
 height12pt depth5.5pt& thin-1 && -1.81 && -78.61 &        \0  & 47.84 &
  -5.974 & -17.65  &\cr
 height12pt depth5.5pt& thin-2 && -4.97 && -50.27 &        \0  & 30.59 &
  -3.820 & -\pdigit 7.46  &\cr
 \noalign{\hrule}
 } } }
}{Thick-wall bubble free energy for \pmb{$\b=1.5$}.}

\puttab{b2.5}{
 \centerline{ \vbox{\offinterlineskip
 \halign to\hsize{\vrule#\tabskip=.5em plus1em minus.5em&
   \hfil# & \vrule# & $#$\hfil & \vrule# &
   $#$\hfil & $#$\hfil & $#$\hfil & $#$\hfil & $#$\hfil
   & \tabskip=0pt\vrule#\cr
 \noalign{\hrule}
 height14pt depth7.5pt& && \hfil\dfo &&\multispan5 \hfil$\dff$\hfil &\cr
 height17pt depth7.5pt& Method && && \hfil T^2 & \hfil T \ln(T) & \hfil T &
   \hfil \ln(T) & \hfil 1 &\cr
 \noalign{\hrule}
 height14pt depth5.5pt&   sub  && -1.34  && -24.90 &        \0  & 17.17 &
  -1.408 & -4.60  &\cr
 height12pt depth5.5pt& pot(A) && -1.009 && -24.90 &        \0  & 14.05 &
  -1.408 & -4.64  &\cr
 height12pt depth5.5pt& der(A) && -1.009 && -24.90 &        \0  & 13.35 &
  -1.408 & -4.64  &\cr
 height12pt depth5.5pt& pot(B) && -1.009 && -24.90 & \hfil 2.48 & 15.60 &
  -1.408 & -4.64  &\cr
 height12pt depth5.5pt& thin-1 && -0.572 && -24.90 &        \0  & 15.15 &
  -1.892 & -5.59  &\cr
 height12pt depth5.5pt& thin-2 && -0.553 && -\pdigit 5.59 & \0  &\pdigit 3.40 &
  -0.424 & -0.83  &\cr
 \noalign{\hrule}
 } } }
}{Thick-wall bubble free energy for \pmb{$\b=2.5$}.}

\sec{Conclusions: A New Prefactor, and Derivative Corrections}

We have tested the effective potential approximation to the critical bubble
free energy.  The agreement is best if one pulls a factor of $\mu^4/T^4$ into
the decay rate prefactor, \puteq{g3}, and takes the real part of the effective
potential in the region $V''<0$ (Method A).  That is, $F_c^{pot(A)}$ closely
approximates $F_c^{sub} \equiv F_c^{trad} - 4T\ln(T/\mu)$.  Table~\putlab{abc}
shows that no single set of $\Ineg$ parameters $\{a,b,c\}$ does consistently
better than Method A.  With scales restored, $E_c = \ord{\sigma^2/\mu}$,
$\dff^{sub} = \ord{T^2/\mu}$, and ``derivative corrections'' are
$$\dff^{sub} - \dff^{pot(A)} = \ord{T} \;.\eq{conc1}
This difference is numerically fairly small, and very poorly predicted by the
derivative expansion [\puteq{fd2}].  In summary,
$${\Gamma\over{\cal V}} = X\; {\mu^4\over2\pi} \left(E_c\over 2\pi T \right)^{3
  /2} {|\o_-|/2T \over \sin(|\o_-|/2T)} \; e^{-F_c^{pot(A)}/T} \;,\eq{conc2}
where $X$ is a dimensionless number representing derivative corrections,
typically $10^{-2}$ to $10^2$.

In 1D, where $\dff^{sub}$ is only $\ord{T}$, derivative corrections [still
$\ord{T}$, and numerically larger] are much more significant than in 3D.

\sectionnumstyle{blank}
\sec{Acknowledgments}
\vskip-.1in
The authors thank Carlos Arag\~ao de Carvalho, Dan Boyanovsky, Rich Holman,
Stephen Hsu, Andrei Linde, Erick Weinberg, and Mark Wise for valuable
discussions.  DB thanks the Aspen Center for Physics for its hospitality during
part of this work.


\sec{References:}
\vskip-.1in
\def\ap#1{{\it Ann.\ Phys.} {\bf #1}}
\def\pl#1{{\it Phys.\ Lett.} {\bf #1}}
\def\np#1{{\it Nucl.\ Phys.} {\bf #1}}
\def\pr#1{{\it Phys.\ Rev.} {\bf #1}}
\def\prl#1{{\it Phys.\ Rev.\ Lett.} {\bf #1}}

\def\rbf#1{{\it Rev.\ Bras.\ Fis.} {\bf #1}}
\def\zp#1{{\it Z.\ Phys.} {\bf #1}}
\def\jcp#1{{\it J.\ Chem.\ Phys.} {\bf #1}}
\def\yf#1{{\it Yad.\ Fiz.} {\bf #1}}
\def\sjnp#1{{\it Sov.\ J.\ Nucl.\ Phys.} {\bf #1}}
\def\ibid{{\it ibid.\ }}

\baselinestretch=1000
\begin{putreferences}
\reference{ejw}{Errors that arise when particles are naively integrated out
  are discussed (at $T=0$) by E. Weinberg, \pr{D47}:4614 (1993).}
\reference{cc}{S. Coleman, \pr{D15}:2929 (1977), {\bf 16}:1248(E) (1977);\\
  C.G. Callan \& S. Coleman, \pr{D16}:1762 (1977);\\
  S. Coleman, ``The Uses of Instantons'', {\it Proc. 1977 Int. School of
   Subnuclear Physics, Ettore Majorana}, ed. A. Zichichi (Plenum, New York,
   1979); reprinted in {\it Aspects of Symmetry} (Cambridge University Press,
   1985).}
\reference{affleck}{I. Affleck, \prl{46}:388 (1981).}
\reference{fTref}{J.W. Cahn \& J.E. Hilliard, \jcp{31}:688 (1959);\\
  J.S. Langer, \ap{41}:108 (1967); \ibid{\bf 54}:258 (1969);\\
  M.B. Voloshin, I.Y. Kobzarev \& L.B. Okun', \yf{20}:1229 (1974)
   [\sjnp{20}:644 (1975)];\\
  P.H. Frampton, \pr{D15}:2922 (1977);\\
  A.D. Linde, \pl{70B}:306 (1977); \ibid{\bf 100B}:37 (1981); \np{B216}:421
   (1983); \ibid{\bf B223}:544 (1983)(E);\\
  O.J.P. \'Eboli \& G.C. Marques, \rbf{16}:147 (1986).}
\reference{gleiser}{M. Gleiser, G.C. Marques \& R.O. Ramos, \pr{D48}:1571
  (1993).}
\reference{dj}{L. Dolan \& R. Jackiw, \pr{D9}:3320 (1974).}
\reference{haber}{H. Haber \& H.A. Weldon, {\it J. Math. Phys.}{\bf 23},
  1852 (1982).}
\reference{us}{C.G. Boyd {\it et al.}, \pr{D48}:4952 (1993).}
\reference{chan}{L.-H. Chan, \prl{54}:1222 (1985); \prl{56}:404(E) (1986).}
\reference{moss}{I. Moss, D. Toms \& A. Wright, \pr{D46}:1671 (1992).}
\reference{thick}{M. Dine \etal, \pr{D46}:550 (1992);
  {\it ibid.}, \pl{283B}:319 (1992).}
\reference{cnl}{T.-P. Cheng \& L.-F. Li, {\it Gauge Theory of Elementary
  Particle Physics\/} (Oxford U. Press, 1984), eq.~(6.133).}
\reference{alpriv}{Andrei Linde, private communication.}
\reference{dod}{S. Dodelson \& B. Gradwohl, \np{B400}:435 (1993).}
\reference{DHN}{R. F. Dashen, B. Hasslacher \& A. Neveu, \pr{D10}:4114
   (1974); \ibid{\bf D10}:4130 (1974);\\
  R. Rajaraman, {\it Solitons and Instantons\/} (North-Holland, 1987), secs.
   5.3--5.4.}
\reference{walls}{I. Ventura, \pr{B24}:2812 (1981);\\
  C. Arag\~ao de Carvalho \etal, \pr{D31}:1411 (1985);\\
  C. Arag\~ao de Carvalho \etal, \np{B265}:45 (1986);\\
  D. Bazeia \etal, \zp{C46}:457 (1990).}
\reference{clee}{C.L.Y. Lee, CALT-68-1903 (Oct. 1993).}

\end{putreferences}

\newpage ~
\vskip 3.5in \centertext{\bf Figure 1:
  The potential \pmb{$V(\phi)$ } for several \pmb{$\kappa$}'s.}
\vskip 3.5in \centertext{\bf Figure 2:
  Thick-wall bubble profiles \pmb{$\p(r)$ } and \pmb{$V''(r)$}.}
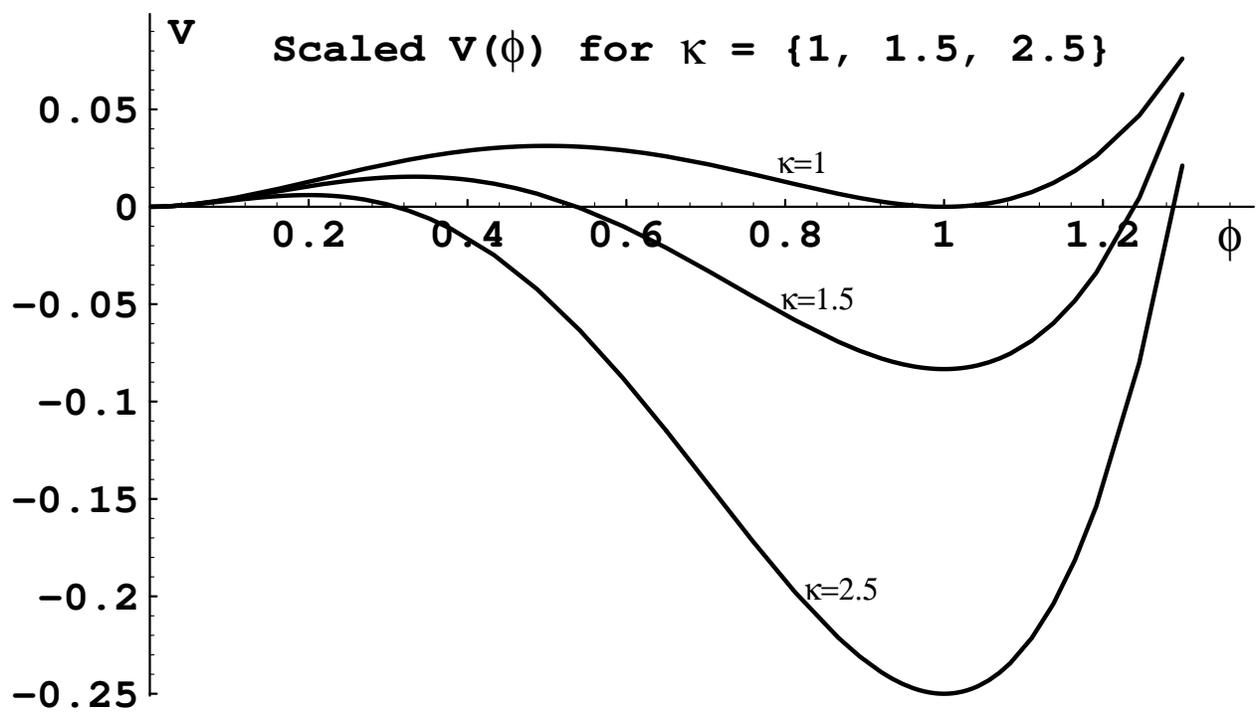
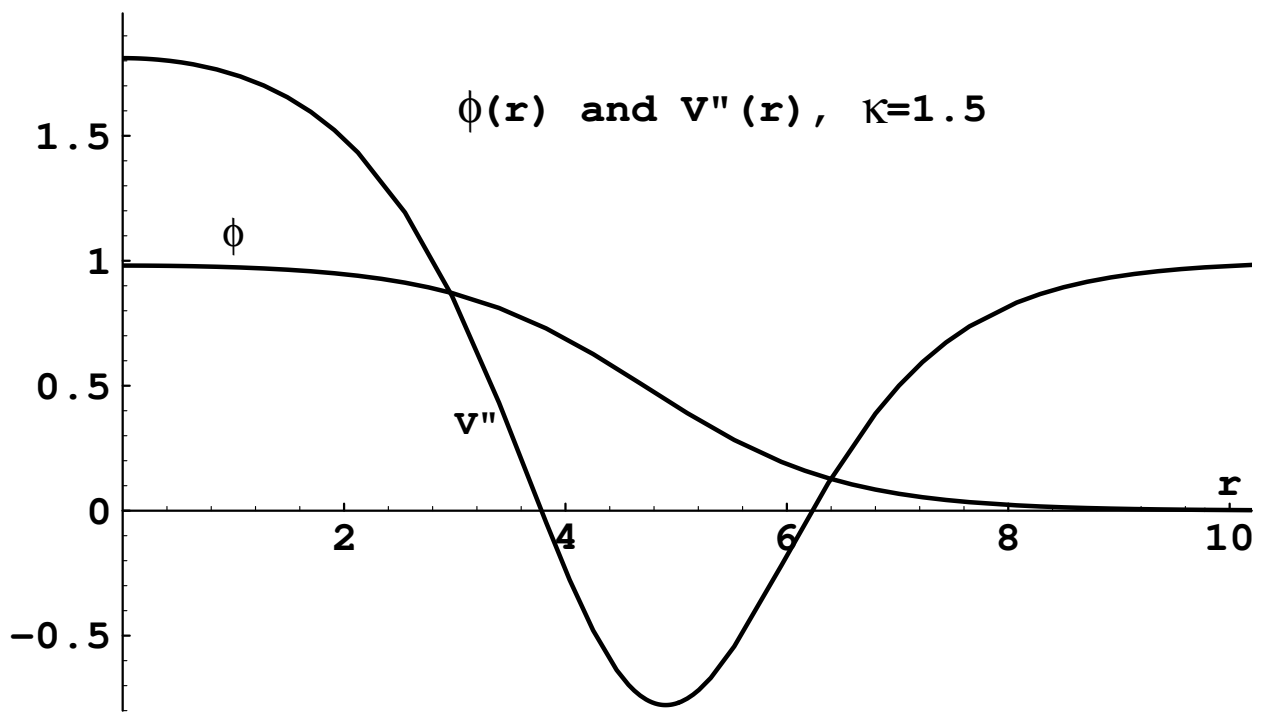
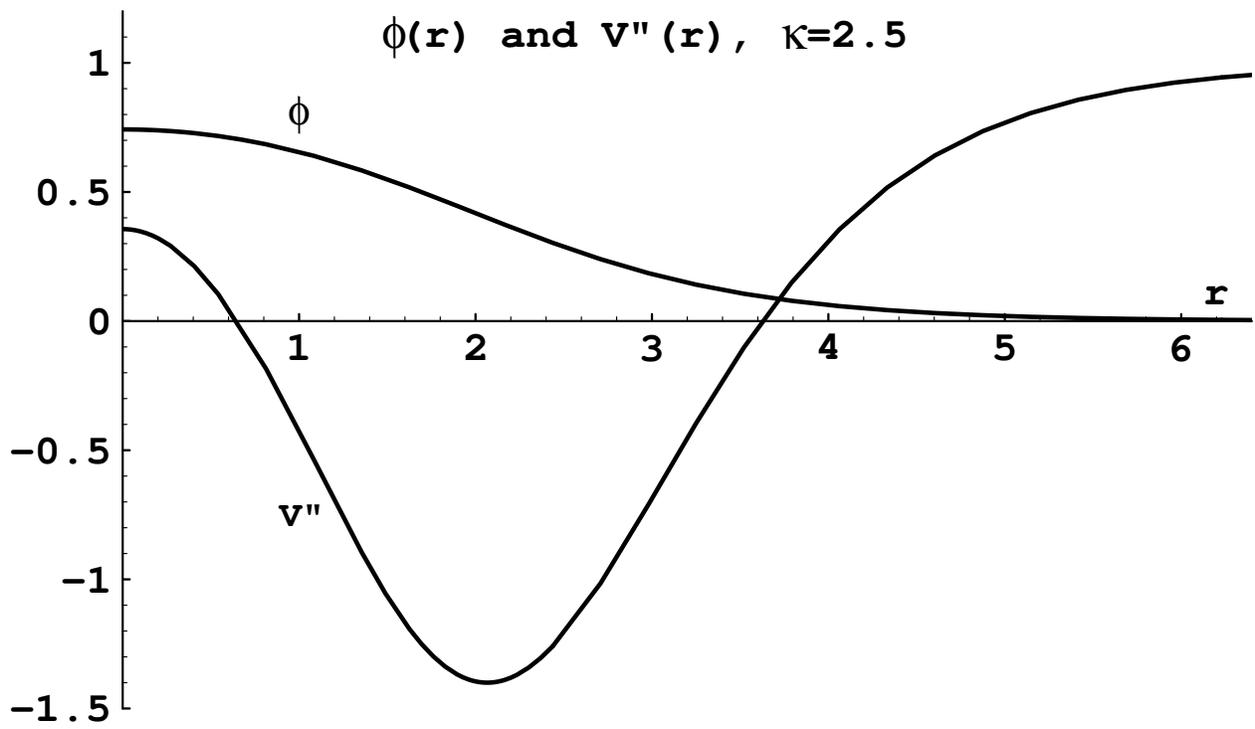
\bye